\documentclass[12pt]{iopart}
\usepackage{graphicx}

\newcommand{\Fe}{\mathcal{G}}
\newcommand{\Fs}{\mathcal{B}}
\newcommand{\anew}{\tilde{a}}
\newcommand{\figwidth}{\columnwidth}

\begin{document}

\title[Universal fluctuation of the one-dimensional KPZ growth]
{Universal fluctuation of the average height in the early-time regime of the one-dimensional Kardar-Parisi-Zhang-type growth}

\author{Deok-Sun Lee}
\address{Theoretische Physik, Universit\"{a}t des Saarlandes, 66041 Saarbr\"{u}cken, Germany}

\author{Doochul Kim}
\address{School of Physics and Astronomy and Center for Theoretical Physics, Seoul National University, 151-747, Korea}

\begin{abstract}
The statistics of the average height fluctuation of the one-dimensional 
Kardar-Parisi-Zhang(KPZ)-type surface is investigated. Guided by the idea of 
local stationarity, we derive the scaling form of the characteristic function 
in the early-time regime, $t\ll N^{3/2}$ with $t$ time and $N$ the system size, 
from the known characteristic function in the stationary state ($t\gg N^{3/2}$) of 
the single-step model derivable from a Bethe Ansatz solution, and thereby find 
the scaling properties of the cumulants and the large deviation function 
in the early-time regime. These results, combined with the scaling analysis of 
the KPZ equation, imply the existence of the universal scaling functions for 
the cumulants and an universal large deviation function. 
The analytic predictions are supported by the simulation results 
for two different models. 
\end{abstract}


\maketitle

\section{Introduction}

The Kardar-Parisi-Zhang (KPZ) equation~\cite{kpz86} is a nonlinear differential
equation with noise representing in the simplest way the roughening phenomena 
of growing surfaces. The roughness is measured by the spatial fluctuation of 
the surface profile, and an initially flat surface of finite size finds its 
width increase in the early-time regime and then saturate in the stationary state. 
The KPZ equation offers a theoretical framework to understand such a kinetic 
roughening, and many real-world nonequilibrium interfaces as well as 
computational models belong to the KPZ class or its variants~\cite{barabasiBook95}. 

A remarkable universal quantity for the (1+1)-dimensional growing surfaces 
is the large deviation function (LDF) \cite{derrida98,derrida99,lee99,appert00}.
It concerns the tail behavior of the probability distribution of 
the spatially-averaged height, $\bar{h}=N^{-1}\sum_{i=1}^N h_i$ 
with $N$ the system size and $h_i$ the height at a position $i$. 
In the stationary state, it has been claimed to be universal within the KPZ class. 
The exact LDF~\cite{derrida98} derived for the single-step 
(SS) model~\cite{plischke87} has the cumulants' ratio $\langle \bar{h}^3\rangle_c^2/
(\langle \bar{h}^2 \rangle_c \langle \bar{h}^4\rangle_c)=0.41517\ldots$, 
which has been reproduced for other computational models with no 
significant deviations~\cite{derrida99,appert00}. The fluctuation of the 
average height from configuration to configuration does not saturate but all of 
its cumulants grow with time even in the stationary state and the LDF has 
asymmetric tails. The LDF of the average height in the stationary state is of 
interest also in the theory of nonequilibrium systems. It has been shown that 
the probability distribution of the phase space contraction (or entropy production) 
rate has a non-trivial symmetry property in dissipative chaotic systems and 
this has been considered as a generalized fluctuation theorem for nonequilibrium 
systems~\cite{gallavotti95}. Recent studies~\cite{lebowitz99} 
show that the theorem is applicable also for stochastic dynamics of interacting 
particles with the symmetry property found in the LDF of the particle current, 
which corresponds to the total height increase, $N\bar{h}$, in the language of 
the growing surfaces. 

The LDF of an averaged quantity contains a lot of information specific 
for a given system. For example, one can obtain the dynamic correlations among 
individual variables, which are not revealed  by the Gaussian behavior 
around the mean value. Previous works, however, have been restricted 
to stationary states, where, in mathematical terms, only the ground state 
of the time-evolution operator survives and thus its LDF could be derived 
analytically. In this work, we explore the LDF of the average height in 
the early-time regime of the KPZ surface. 
By the early-time regime, we mean $N\to\infty$, $t\to\infty$, and $t\ll N^{z}$, with 
$z=3/2$ the dynamic exponent.
In the early-time regime, a flat surface evolves into its stationary state 
which is rough. The LDF of the average height corresponds to the Legendre 
transformation of the logarithm of the characteristic function, as the entropy 
to the free energy. We derive the characteristic function in the early-time regime 
of the SS model from its stationary-state counterpart that is 
exactly known~\cite{derrida98,derrida99,lee99}. To do so, we use 
the assumption that the scaling property and the model-parameter dependence of 
the latter are the same as those of the locally-stationary segments in the early-time 
regime, the size of each of which scales with time as $t^{1/z}$.
Combined with this result, the scaling analysis of the KPZ equation enables us to 
see how the characteristic function depends on the parameters of the KPZ equation. 
We derive the LDF and the dynamic scaling behavior of the cumulants from this 
characteristic function, and their universal parts are identified in the numerical 
results from two different models in the KPZ class.

\section{LDF in the stationary state of the SS model}

The (1+1)-dimensional KPZ equation describes the evolution of the heights 
$h(x,t)$ on a substrate $\{x|0<x\leq L\}$:
\begin{equation}
\frac{\partial h(x,t)}{\partial t} = \nu \frac{\partial^2 h}{\partial x^2} 
+\frac{\lambda}{2}\left(\frac{\partial h}{\partial x}\right)^2 +\eta(x,t),
\label{eq:KPZ}
\end{equation}
where  the noise $\eta(x,t)$ satisfies $\langle \eta(x,t)\eta(x',t')\rangle 
=2D \delta(x-x')\delta(t-t')$~\cite{kpz86}. It is well known that the rescaling of the 
space $x\to bx$ gives rise to $h\to b^\alpha h$ and $t\to b^z t$ with $\alpha=1/2$ and 
$z=3/2$ as a result of the Galilean invariance and the fluctuation-dissipation 
theorem valid in one space dimension~\cite{halpinhealy95}.  

We first review known results about the LDF derived from the Bethe Ansatz solution. 
The SS model~\cite{plischke87} is a discrete growth model belonging to the KPZ class, 
where the height $h_i$ for $1\leq i\leq N$ can take even  (odd) integer numbers 
for $i$ even (odd) and satisfy the SS condition, $|h_{i+1}-h_i|=1$, for all $i$
with $h_{N+1} = h_1$. The height at a randomly chosen site is increased by $2$ 
with probability $(1+\epsilon)/2$ or decreased by $2$ with probability 
$(1-\epsilon)/2$, with $0<\epsilon \leq 1$, provided the 
new height does not violate the SS condition, and the time $t$ increases by 
$1$ after $N$ such attempts. This model is equivalent to the partially 
asymmetric exclusion process. Hence the time-evolution operator for the height 
step configuration $\{\sigma\}$ with $\sigma_i = h_i-h_{i-1}$ is identical  
to the asymmetric XXZ spin chain Hamiltonian which in turn is diagonalized by the 
Bethe Ansatz~\cite{gwa92prl,kim95}. 
The ground state energy of a parameterized version of the Hamiltonian then determines 
the characteristic function of the total height increase, 
$J=\sum_i (h_i(t)-h_i(0))$,  for $tN^{-3/2}\gg 1$. 
Note that the excitation gap scales as $N^{-3/2}$ ($z=3/2$) \cite{gwa92prl,kim95}.
With $\langle \ldots \rangle$ denoting the expectation 
with respect to the distribution of $J$, $P(J)$, the characteristic function of $J$ 
is given by \cite{derrida98,derrida99,lee99}
\[
\langle e^{\gamma J}\rangle = e^{\gamma \langle J\rangle+\Fe(\gamma,N,t)} \ {\rm with}
\]
\begin{equation}
\Fe(\gamma,N,t) =  \frac{\epsilon \sqrt{4\rho(1-\rho)}}{2\sqrt{2\pi}} 
\frac{t}{N^{3/2}}
G\left(
\sqrt{2\pi}\gamma \sqrt{4\rho(1-\rho)}N^{3/2}\right),
\label{eq:Fe1}
\end{equation}
for $\gamma N^{3/2}$ finite, where  
$\langle J\rangle = 2 \epsilon \rho(1-\rho) N^2 t(N-1)^{-1}$ and 
the function $G(x)$ is given parametrically as 
$G = \sum_{n=1}^\infty (-z)^n (n^{-3/2}-n^{-5/2})$ and 
$x=-\sum_{n=1}^\infty (-z)^n n^{-3/2}$ for $x\geq -1$ along with 
its analytic continuation for the region $x<-1$~\cite{lee99}.
From now on, $h_i(t)$  refers to $h_i(t)-h_i(0)$ and $\bar{h}=J/N$.
Then the characteristic function of the average height distribution is simply, 
$\langle e^{\gamma \bar{h}}\rangle 
 = e^{\gamma \langle \bar{h}\rangle + \Fe(\gamma/N,N,t)}$ 
with 
$\langle \bar{h}\rangle = \langle J\rangle/N$.

The logarithm of the average height distribution, $\ln P(\bar{h})$, is related 
to $\ln \langle e^{\gamma \bar{h}}\rangle$ via the Legendre transformation: 
\begin{equation}
\ln P (\bar{h}) = \ln \langle e^{\gamma\bar{h}}\rangle -\gamma \bar{h} \quad {\rm and} \quad
\bar{h} = \frac{\partial}{\partial \gamma}\ln \langle e^{\gamma\bar{h}}\rangle .
\label{eq:Legendre}
\end{equation}
Using equation (\ref{eq:Fe1}), one finds that for $t\gg N^{3/2}$, 
\[
P(\bar{h}=\langle \bar{h}\rangle+yN/t)\sim e^{-tf(y)} \ {\rm with} 
\]
\begin{equation}
f(y)=\frac{1}{N^{3/2}} \frac{\epsilon \sqrt{4\rho(1-\rho)}}{2\sqrt{\pi}} H\left(
    \frac{y }{2\epsilon \rho(1-\rho)}\right),
\label{eq:ldf1}
\end{equation}
where the function $H(x)$ is defined parametrically as $H=[G(z) - z(\partial/\partial z)G(z)]/\sqrt{2}$ and $x=(\partial/\partial z)G(z)$ 
and behaves as $H(x)\sim |x|^{3/2}$ for $x\to-\infty$ and $H(x)\sim x^{5/2}$ 
for $x\to\infty$~\cite{lee99}. 

\begin{figure}
\begin{center}
\includegraphics[width=\figwidth]{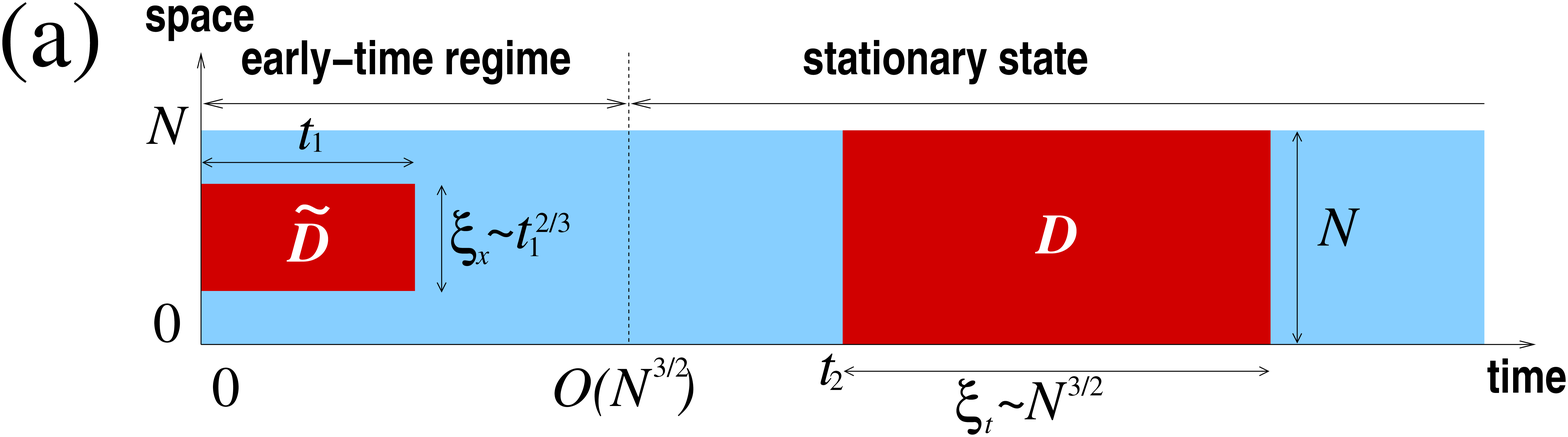}
\includegraphics[width=\figwidth]{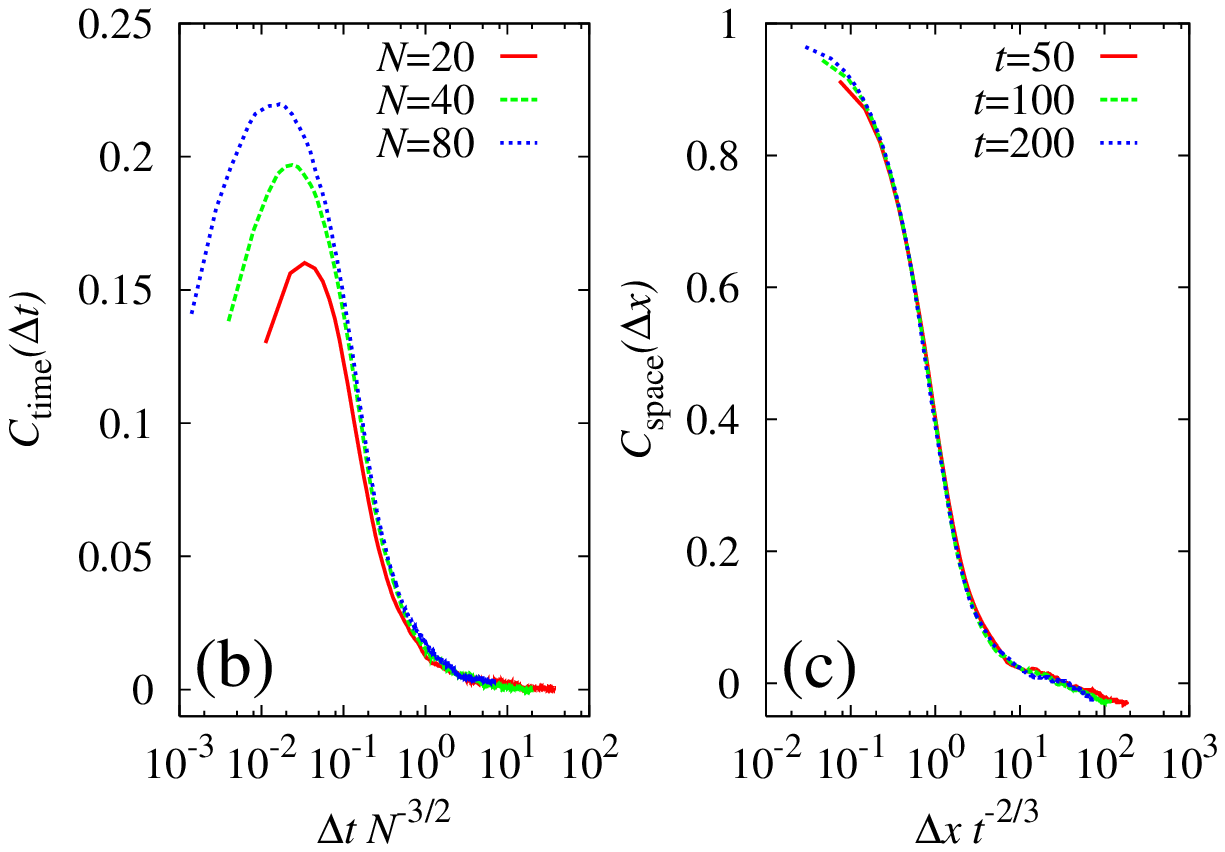}
\caption{Correlation of the partial height increases. 
(a) Example of a stationary correlated domain in the stationary state, $D$, and 
  that in the early-time regime, $\tilde{D}$, on both of which 
  all the partial height increases are correlated and the height-steps are uncorrelated.
(b) Plot of $C_{\rm time}(\Delta t)$ versus $\Delta t N^{-3/2}$ for the SS model with $\epsilon=1$ and 
$\rho=1/2$, and the system size $N=20, 40$, and $80$.  (c) Plot of 
$C_{\rm space}(\Delta x)$ versus $\Delta x t^{-2/3}$ for the SS model with 
$\epsilon=1, \rho=1/2, N=10^4$ at time  $t=50, 100$, and $200$.} 
\label{fig:correlation}
\end{center}
\end{figure}

\section{LDF in the early-time regime of the SS model}

\subsection{Local stationarity}

A feature of the stationary state of the one-dimensional KPZ surface 
is that the height-steps at different sites 
and at equal or different times are uncorrelated as the fluctuation-dissipation 
theorem states~\cite{halpinhealy95}.
On the other hand, the height increases, $\bar{h}(t+\Delta t)-\bar{h}(t)$ with 
different $t$'s may be correlated. For example, a flat surface has many sites 
where the heights can increase while in a $V$-shaped surface the heights has  
only a few such sites. Moreover, such morphological features relevant to the height 
increase can survive over a certain period of time $\xi_t$. 
The factor $t/N^{3/2}$ in $\Fe(\gamma,N,t)$ implies that $\xi_t\sim N^{3/2}$ 
[figure~\ref{fig:correlation}(a)]. 
This motivates us to decompose the total height increase $J$ into 
$t/N^{3/2}$ effectively-independent partial height increases, as 
$J =\sum_{i=0}^{t/N^{3/2}} J_{iN^{3/2},0} (N^{3/2},N)$, 
where 
\begin{equation} 
J_{t,x}(\Delta t, \Delta x) \equiv  \sum_{i=x+1}^{x+\Delta x} [h_i(t+\Delta t)-h_i(t)] .
\end{equation}  
In this notation, $J=J_{0,0}(t,N)$. 
That such partial height increases are statistically independent is our main  
assumption. To check its validity we calculate, for the SS model,    
the normalized correlation defined as 
\begin{equation}
C_{\rm time}(\Delta t)= \frac{\langle J_{t,0}(\Delta t,N)\Delta J_{t+\Delta t,0}(\Delta t,N)\rangle_c}
{\sqrt{\langle J_{t,0}(\Delta t,N)^2\rangle_c \langle J_{t+\Delta t,0}(\Delta t,N)^2\rangle_c}}
\end{equation} 
around $\Delta t\sim N^{3/2}$ 
where $\langle AB \rangle_c = \langle (A-\langle A\rangle)(B-\langle B\rangle)\rangle$. 
The results shown in figure~\ref{fig:correlation}(b) exhibit a rapid decay of the 
correlation for $\Delta t \gg N^{3/2}$ and confirm the mutual independence 
of $J_{iN^{3/2},0}(N^{3/2},N)$ with different $i$'s.

In the early-time regime when the roughening is still ongoing, the height-steps  
at time $t$ are uncorrelated only within each segment of size $\xi_x\sim t^{2/3}$, 
and such locally-stationary segments expand as time goes on 
and merge into one in the stationary state. 
At a given time $t_1$ with $1 \ll t_1 \ll N^{3/2}$, we may assume that 
most of the height steps are uncorrelated and all of the partial height 
increases are correlated within a domain 
$[0,t_1]\times [x,x+\xi_x]$  with $\xi_x\sim t_1^{2/3}$ 
($\tilde{D}$ in figure~\ref{fig:correlation}(a)),  
in the same way as within a stationary correlated domain 
$[t_2,t_2+\xi_t]\times [0,N]$ of the stationary state with $\xi_t\sim N^{3/2}$ and 
$t_2\gg N^{3/2}$ ($D$ in figure~\ref{fig:correlation}(a)). 
To check this idea, we measure for the SS model the 
normalized correlation function, 
\begin{equation} 
C_{\rm space}(\Delta x) 
= \frac{\langle J_{0,x}(t,\Delta x) J_{0,x+\Delta x}(t,\Delta x)\rangle_c}{\sqrt{\langle 
  J_{0,x}(t,\Delta x)^2 \rangle_c \langle J_{0,x+\Delta x}(t,\Delta x)^2\rangle_c}},
\end{equation} 
as shown in figure~\ref{fig:correlation}(c). 
The decaying behavior of $C_{\rm space}(\Delta x)$  confirms  the mutual independence 
of the partial height increases, 
$J_{0,i\xi_x}(t,\xi_x)$ and $J_{0,j\xi_x}(t,\xi_x)$ 
on different ($i\ne j$) domains of size $t\times \xi_x$.

\subsection{Dynamic scaling and LDF in the early-time regime of the SS model} 

The same correlation landscape between the stationary correlated domains 
of the stationary state and of the early-time regime implies  
that the characteristic function in the stationary state, 
$e^{\gamma \langle J\rangle+\Fe(\gamma,N,t)}$ 
gives the characteristic function for a stationary correlated domain of 
the early-time regime, $\langle e^{\gamma J_{0,x}(t,\xi_x)}\rangle$, 
provided the system size $N$ in the former is replaced by the correlation 
length $\xi_x\sim \anew t^{2/3}$ in the latter with $\anew$ a constant. 
In this mapping, actual  
exact functional form of $G(x)$ in equation (2) may possibly change.
Then the characteristic function for the early-time height increase  
is determined as 
\begin{equation} 
\langle e^{\gamma J}\rangle=\langle e^{\gamma \sum_{i=0}^{N/\xi_x} J_{0,i\xi_x}(t,\xi_x)}\rangle 
\sim \langle e^{\gamma J_{0,x}(t,\xi_x)}\rangle^{N/\xi_x} .
\end{equation}
We then have, with $t\ll N^{3/2}$,
 \[
 \langle e^{\gamma J}\rangle = e^{\gamma\langle J\rangle + \tilde{\Fe}(\gamma,N,t)} \ 
{\rm with}
\]
\begin{equation}
\tilde{\Fe}(\gamma,N,t)= \frac{N}{\anew t^{2/3}} \frac{1}{{\anew}^{3/2}}
\frac{\epsilon \sqrt{4\rho(1-\rho)}}{2\sqrt{2\pi}}
    \times \tilde{G}\left(
\sqrt{2\pi}\gamma \sqrt{4\rho(1-\rho)}{\anew}^{3/2} t\right)
\label{eq:Fe2}
\end{equation}
for $\gamma t$ finite with a new function $\tilde{G}(x)$ introduced. 

\begin{figure}
\begin{center}
\includegraphics[width=\figwidth]{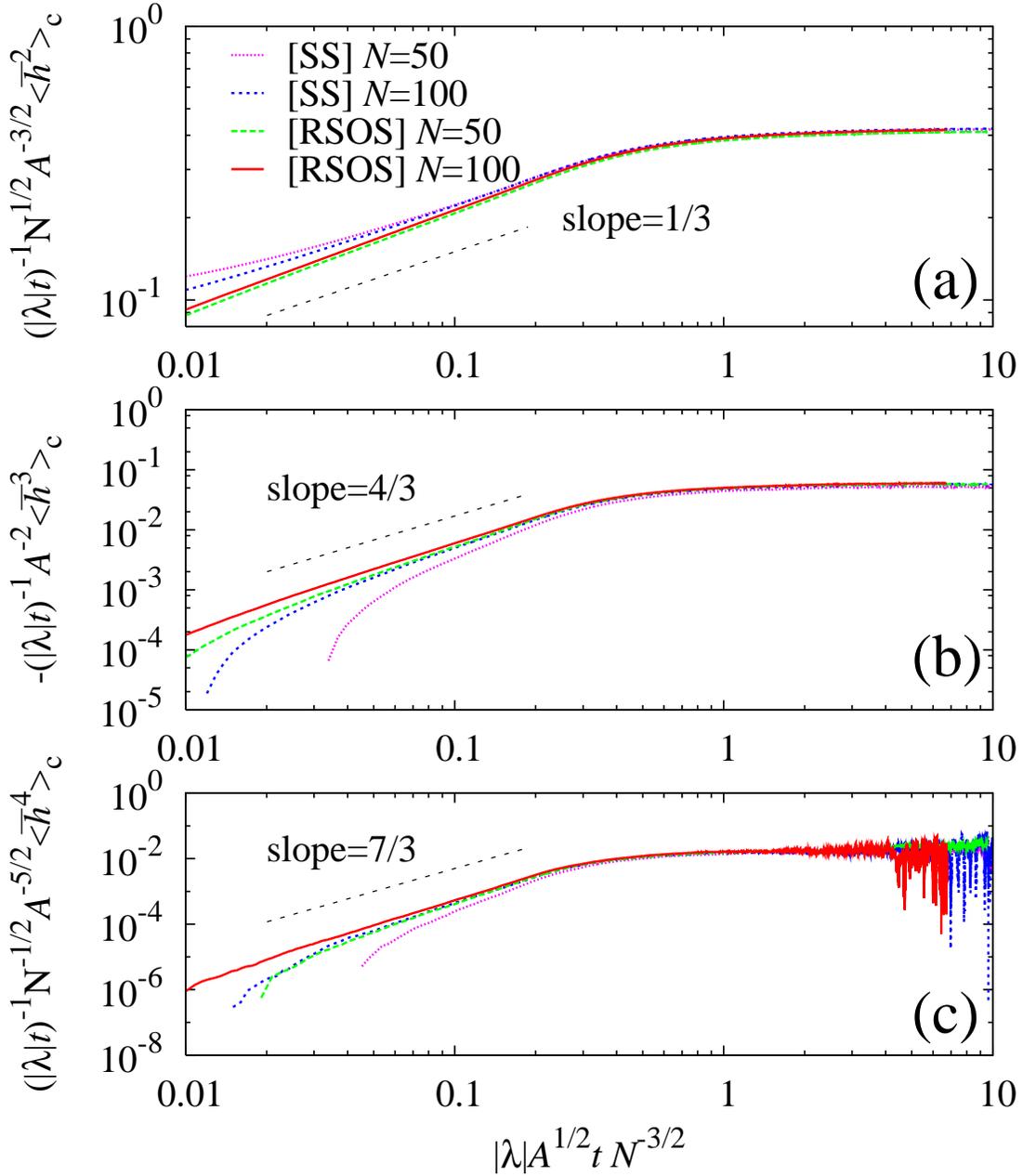}
\caption{Dynamic scaling of the (a) second, (b) third, and (c) fourth cumulants of the average height 
of the SS model with $\epsilon=1$ and $\rho=1/2$ ($\lambda=-1, A=1$) and the 
  RSOS model~\cite{kim89} ($\lambda=-0.75, A=0.81$) 
  for $N=50$ and $100$. The data scaled according to 
equations~(\ref{eq:cumulants}) and (\ref{eq:cumulants2}) are presented, 
showing good collapse and also confirming the small-$x$ behavior 
  of the scaling functions $\phi_n(x)$. The legend is provided in (a). }
\label{fig:cumulants}
\end{center}
\end{figure}

The expressions of $\Fe(\gamma,N,t)$ and $\tilde{\Fe}(\gamma,N,t)$ 
together determine the dynamic scaling of the average height 
by the relation 
$\langle \bar{h}^n\rangle_c = N^{-n} (\partial^n /\partial \gamma^n) 
\ln \langle e^{\gamma J}\rangle|_{\gamma=0}$, 
which yields for $n=2, 3, \ldots$,  
\[
\langle \bar{h}^n\rangle_c = 
tN^{(n-3)/2} \epsilon [4\rho(1-\rho)]^{(n+1)/2} 
\phi_n\left(\frac{t}{(N/\anew)^{3/2}}\right) \ {\rm with} 
\]
\begin{equation}
\phi_n(x)\simeq \left\{
\begin{array}{lc}
2^{(n-3)/2}\pi^{(n-1)/2} G^{(n)}(0) & (x\gg 1), \\
 2^{(n-3)/2}\pi^{(n-1)/2}\tilde{G}^{(n)}(0) \  x^{n-5/3} & (x\ll 1).
\end{array}
\right.
\label{eq:cumulants}
\end{equation}
Here $G^{(n)}(0) = (\partial^n/\partial z^n)G(z)|_{z=0}$ and 
the same for $\tilde{G}^{(n)}(0)$.
The scaled data of the $2$nd, $3$rd, and $4$th cumulants  in the SS model 
with $\epsilon=1$ and $\rho=1/2$ are plotted in figure~\ref{fig:cumulants}, 
which agree with equation~(\ref{eq:cumulants}).

From equation~(\ref{eq:Fe2}) one also finds that 
the LDF for $N\gg t^{2/3}$ is given by 
\[
P(\bar{h}=\langle \bar{h}\rangle + \tilde{y}/t^{1/3}) \sim e^{-N\tilde{f}(\tilde{y})} \ {\rm with} 
\]
\begin{equation}
\tilde{f}(\tilde{y})=\frac{1}{t^{2/3}} 
\frac{1}{{\anew}^{5/2}}\frac{\epsilon \sqrt{4\rho(1-\rho)}}{2\sqrt{\pi}} \tilde{H}\left(
    \frac{\tilde{y}\tilde{a}}{2\epsilon \rho(1-\rho)}\right),
\label{eq:ldf2}
\end{equation}
where $\tilde{H}(x)$ is defined by 
$\tilde{H}=[\tilde{G}(z) - z(\partial/\partial z)\tilde{G}(z)]/\sqrt{2}$ and 
$x=(\partial/\partial z)\tilde{G}(z)$. 

\section{Scaling analysis and universal LDF} 

The LDF and the dynamic scaling behaviors shown in equations~(\ref{eq:cumulants}) 
and (\ref{eq:ldf2}) can be represented in terms of the parameters 
of the KPZ equation (\ref{eq:KPZ}), which enables us to check 
their universality. Following the procedures proposed in Ref.~\cite{amar92b}, 
one finds the following dimensionless variables
\begin{equation}
h' = \frac{\lambda}{2\nu} h, \ \ 
x'=\frac{\lambda^2 D}{2\nu^3} x, \ \
\tau = \frac{\lambda^4 D^2}{4\nu^5} t 
\end{equation}
satisfy an equation without parameters, 
$(\partial /\partial \tau)h' = (\partial^2/\partial x^2)h' 
+(\partial h/\partial x)^2 +\eta'(x',\tau)$ with 
$\langle \eta'(x_1',\tau_1)\eta'(x_2',\tau_2)\rangle 
=\delta(x_1'-x_2')\delta(\tau_1-\tau_2)$. 
Therefore both $\Fe$ and $\tilde{\Fe}$ should be functions of 
$\gamma N (\lambda/\nu)$ (for $\gamma J$), $ (\lambda^2 D/\nu^3)N$, and $(\lambda^4 D^2/\nu^5)t$. 
Keeping the $t$ and $N$  dependence appearing in equations~(\ref{eq:Fe1}) and (\ref{eq:Fe2}), 
those functions turn out to be given by 
\begin{equation}
\Fe(\gamma,N,t) = |\lambda|\sqrt{A} \frac{t}{N^{3/2}} 
\Fs\left({\rm sign}(\lambda) \sqrt{A} \gamma\, N^{3/2}\right),
\label{eq:Fe1scal}
\end{equation}
where $A\equiv D/\nu$, and 
\begin{equation}
\tilde{\Fe}(\gamma,N,t) = \frac{1}{|\lambda|^{2/3} A^{1/3}} \frac{N}{t^{2/3}} 
\tilde{\Fs}\left(\lambda A \gamma \, t\right).
\label{eq:Fe2scal}
\end{equation}
Here, $\Fs (x)$ and $\tilde{\Fs} (x)$ are the scaling functions 
to be identified as follows.
The nonlinear term with the parameter $\lambda$ in equation~(\ref{eq:KPZ})
has been introduced to describe the lateral growth and thus 
$\lambda = (\partial^2/\partial u^2)(\partial /\partial t) \langle \bar{h}\rangle|_{u=0}$ 
with $u$ the average slope of the surface~\cite{krug92}, which 
leads to $\lambda=-\epsilon$ for the SS model where 
$\langle \bar{h}\rangle = 2\epsilon\rho(1-\rho)t$  and $u = 1-2\rho$. 
Also, the uncorrelated height-steps in the stationary state 
result in $\langle (h_i(t)-h_j(t))^2\rangle_c\sim A|i-j|$~\cite{huse85}, which 
gives   $A=4\rho(1-\rho)$ from the comparison with the exact solution for the SS 
model~\cite{derrida97}. Substituting these values to equation~(\ref{eq:Fe1scal}), one 
finds that equations~(\ref{eq:Fe1}) and (\ref{eq:Fe1scal}) perfectly agree with 
each other under the relation $\Fs(x)=G(-\sqrt{2\pi}x)/(2\sqrt{2\pi})$. Also 
equations~(\ref{eq:Fe2}) and (\ref{eq:Fe2scal}) become equal under the relation 
$\anew  = |\lambda|^{2/3}A^{1/3} \tilde{c}$ with $\tilde{c}$ an universal 
constant and the identity 
$\tilde{\Fs}(x)= \tilde{G}(-\sqrt{2\pi}x\tilde{c}^{3/2}) /(2\sqrt{2\pi}\tilde{c}^{5/2})$. 

With the functions $\Fe$ and $\tilde{\Fe}$ in equations (\ref{eq:Fe1scal}) and 
(\ref{eq:Fe2scal}), we find
\begin{equation}
\langle \bar{h}^n\rangle_c = 
tN^{(n-3)/2} |\lambda| A^{(n+1)/2} 
\phi_n\left(|\lambda| A^{1/2}\frac{t}{N^{3/2}}\tilde{c}^{3/2}\right).
\label{eq:cumulants2}
\end{equation}
To check the $\lambda$ and $A$ dependence in equation~(\ref{eq:cumulants2}) 
and thereby explore the universal scaling function $\phi_n(x)$, we 
performed numerical simulations of the RSOS model~\cite{kim89}, where the flat 
surface, $h_i=0$ for all $i$ at $t=0$, grows via the integer-valued height 
at a randomly-chosen site being increased by $1$ with probability $1$ unless 
the new height differs from neighboring heights by larger than $1$. The numerical 
estimates have shown that $\lambda\simeq -0.75$ and $A\simeq 0.81$ for the RSOS 
model~\cite{krug92}. Using these values, we plot the scaled data of the $n$th 
cumulants with $n=2, 3,$ and $4$ obtained from the RSOS model together with those 
from the SS model ($\epsilon=1$ and $\rho=1/2$) in figure~\ref{fig:cumulants}. 
The good collapse of those data obtained for different times and  
system sizes strongly supports the existence of the universal scaling function $\phi_n(x)$.

\begin{figure}
\begin{center}
\includegraphics[width=\figwidth]{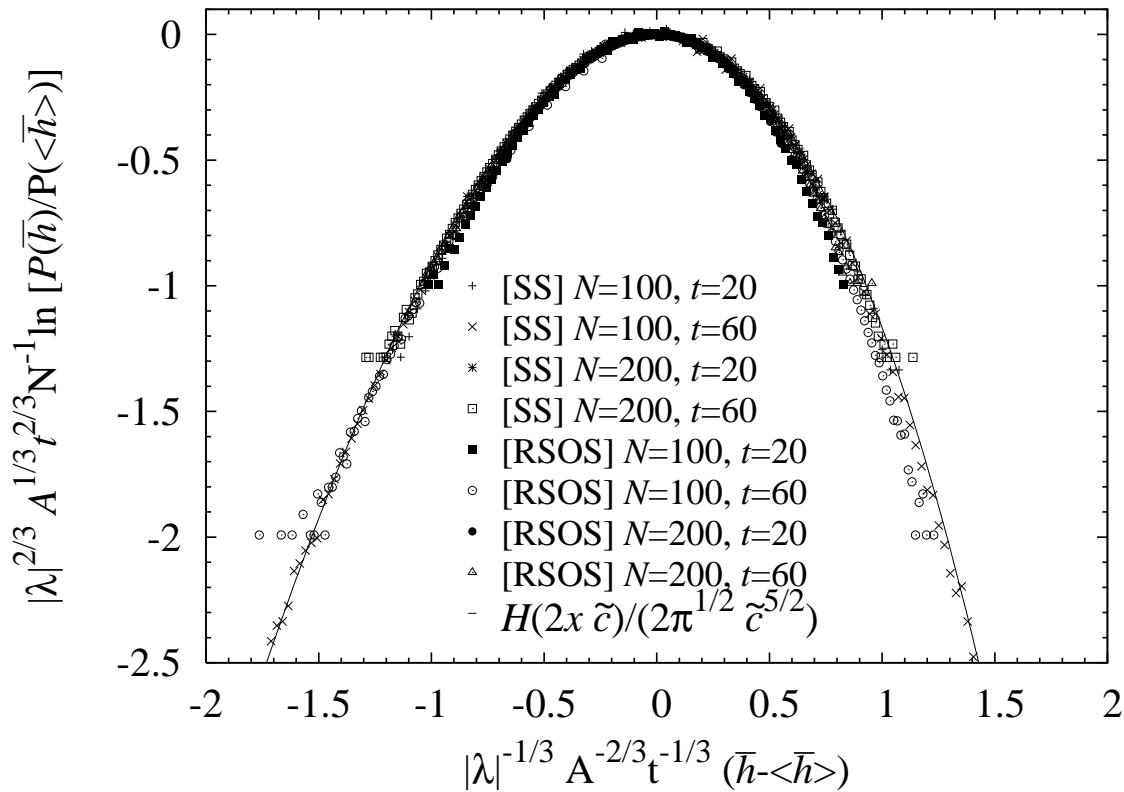}
\caption{Universal LDF in the early-time regime. The data scaled according to 
equation~(\ref{eq:ldf2scal}) are shown, obtained from the simulations of 
the SS model ($\epsilon=1$ and $\rho=1/2$) and the RSOS model with $N=100$ and $200$ and $t=20$ and $60$. 
The collapsed data show the universal function 
$(2\sqrt{\pi}\tilde{c}^{5/2})^{-1}\tilde{H}(2\tilde{c}x)$ in equation
(\ref{eq:ldf2scal}). To compare the functions 
$H(x)$ and $\tilde{H}(x)$, we draw the curve representing the function 
$H(2x\tilde{c})/(2\pi^{1/2}\tilde{c}^{5/2})$ with $\tilde{c}=1.2$ chosen 
for the best fitting.}
\label{fig:uni_ldf}
\end{center}
\end{figure}

The Legendre transformations of equations~(\ref{eq:Fe1scal}) and (\ref{eq:Fe2scal}) 
give the LDF as 
\begin{equation}
f(y)=\frac{1}{N^{3/2}}|\lambda| \sqrt{A} 
\frac{1}{2\sqrt{\pi}}H\left( -2\frac{y}{\lambda A}\right)
\end{equation}
for the stationary state and  
\begin{equation}
\tilde{f}(\tilde{y})=\frac{1}{t^{2/3}}\frac{1}{|\lambda|^{2/3} A^{1/3}} \frac{1}{2\sqrt{\pi}
\tilde{c}^{5/2}}
\tilde{H}
\left( -2\tilde{c} \, {\rm sign}(\lambda) \frac{\tilde{y}}{|\lambda|^{1/3} A^{2/3}}\right)
\label{eq:ldf2scal}
\end{equation}
for the early-time regime, respectively.
While the acquisition of the data for the LDF of the stationary state is hampered by 
long evolution times, that of the early-time regime 
can be obtained within reasonable times.   
We plot $|\lambda|^{2/3}A^{1/3}t^{2/3}N^{-1}\ln [P(\bar{h})/P(\langle \bar{h}\rangle)]$ 
versus $|\lambda|^{-1/3}A^{-2/3}t^{-1/3}(\bar{h}-\langle \bar{h}\rangle)$ with different 
system sizes and times in the SS model and the RSOS model  in figure~\ref{fig:uni_ldf}, 
which shows clearly a universal curve $(2\sqrt{\pi}\tilde{c}^{5/2})^{-1}
\tilde{H}(2\tilde{c}x)$. Note that $\lambda$ is negative for both models.

Contrary to the stationary state, the exact functional form of the universal 
LDF of the early-time regime is not known yet since all the eigenstates of 
the time-evolution operator are involved in the characteristic function. 
From the physics viewpoint, the detailed functional form of the probability 
distribution of the sum of correlated variables is expected to depend 
not only on the type of correlation but also  on the type of the boundary 
condition~\cite{appert00}. Since the exact universal LDF 
of the stationary state has been obtained under the periodic boundary 
condition, 
it is not possible to draw the conclusion that 
$H(x)=\tilde{H}(x)$ on the analytical basis. 
The boundary condition for the locally-stationary segment in the early-time 
regime is not obvious. On the one hand, however, the functions $H(x)$ and 
$\tilde{H}(x)$ 
are expected to display similar behaviors due to the same type of correlations 
among height increases and as seen in figure~\ref{fig:uni_ldf}, the function 
$\tilde{H}(x)$ has indeed asymmetric tails as $H(x)$ does. 
Moreover, we compare $H(x)$ and $\tilde{H}(x)$ in figure~\ref{fig:uni_ldf} 
by drawing the curve representing the function 
$(2\sqrt{\pi}\tilde{c}^{5/2})^{-1}H(2x\tilde{c})$ with $\tilde{c}=1.2$ chosen 
for the best fitting. The analytic curve and the numerical one are 
in good agreement over the whole range of data presented. This 
is consistent with a conjecture that $H(x)$ = $\tilde{H}(x)$.

\section{Conclusion}
We have investigated the statistics of the average height 
for the one-dimensional KPZ surface. 
The LDF in the stationary state implies that 
the average height is decomposed into statistically-independent 
components each of which is a set of correlated variables, and we 
checked its validity.
To derive the LDF in the early-time regime, 
we assumed that  the statistics of the average height in such a component remains 
the same across both time regimes, in that the scaling property and 
the model-parameter dependence of the characteristic function are not changed.
The scaling analysis of the KPZ equation enables us to find 
the equation-parameter dependence of the LDF. Using this  result 
in analyzing the simulation data obtained for two different models, 
we confirmed the universality of the LDF.

We combined the exact results and the scaling analysis to explore 
the universality of the LDF, which can be complementary to 
the measurement of the cumulants' ratio. 
The exact functional form of the universal LDF is an open question.
Given that little is known about the statistical properties of non-stationary 
systems, the present work could stimulate related research including the 
investigation of the fluctuations in non-stationary systems.

\ack

We thank useful discussions with H. Park, J.M. Kim, and Y. Kim. 
This work is supported by the KRF Grant No. R14-2002-059-010000-0 
in the ABRL program funded by the Korean government MOEHRD.

\end{document}